\documentclass[floatfix, a4paper, amssymb, amsmath, reprint, aps, nofootinbib, twoside, pra]{revtex4-2}
\usepackage[utf8]{inputenc}
\usepackage{graphicx}
\usepackage{physics}
\usepackage{hyperref}
\usepackage{xcolor}
\usepackage[draft]{changes}
\definechangesauthor[name={Rajibul}, color=violet]{RI}
\definechangesauthor[name={Zewen}, color=red]{zs}
\definechangesauthor[name={Yi Hong}, color=blue]{yh}

\begin{document}


\title{Investigations of 2D ion crystals in a hybrid optical cavity trap for quantum information processing}
\author{Zewen Sun$^1$, Yi Hong Teoh$^1$, Fereshteh Rajabi$^1$, Rajibul Islam$^1$}
\affiliation{$^1$Institute for Quantum Computing and Department of Physics and Astronomy, University of Waterloo, 200 University Ave. West, Waterloo, Ontario N2L 3G1, Canada}


\begin{abstract}
We numerically investigate a hybrid trapping architecture for 2D ion crystals using static electrode voltages and optical cavity fields for in-plane and out-of-plane confinements, respectively. 
By studying the stability of 2D crystals against 2D-3D structural phase transitions, we identify the necessary trapping parameters for ytterbium ions. 
Multiple equilibrium configurations for 2D crystals are possible, and we analyze their stability by estimating potential barriers between them.
We find that scattering to anti-trapping states limits the trapping lifetime, which is consistent with recent experiments employing other optical trapping architectures.  
These 2D ion crystals offer an excellent platform for quantum simulation of frustrated spin systems, benefiting from their 2D triangular lattice structure and phonon-mediated spin-spin interactions. 
Quantum information processing with tens of ions is feasible in this scheme with current technologies.
\end{abstract}

\maketitle


\section{Introduction}
\label{sec:intro}
Two-dimensional controlled many-body quantum systems, with their enriched phase diagrams, open up a new realm for physics study that are not readily accessible in one-dimension, e.g.~geometric magnetic frustration \cite{diep2013frustrated, moessner2006geometrical} and topological order \cite{wen1991mean,savary2016quantum}. 
Among all versatile quantum simulators, trapped ion systems are an excellent platform to investigate quantum information processing (QIP) experiments \cite{Monroe2021programmable,Blatt2012quantum}. 
Most conventional ion traps employ radio-frequency (RF) fields in addition to static (DC) potentials to create ion confinement.
Typical geometrical constraints of RF-trap electrodes allow trapping of a chain of ions whose equilibrium positions are made to coincide with the `RF-null' line.
This is to avoid the so-called `micromotion heating' problem where the driving RF fields cause unwanted heating of vibrational modes of ions that are used to mediate quantum entanglement. 
Despite recent efforts in developing RF traps with microfabricated electrodes for an ensemble of individual ions in a flexible geometry \cite{Seidelin2006microfabricated, Schmied2009optimal}, creating two-dimensional micromotion-free arrangement of ions remains a challenging technical task.

Two-dimensional ion systems for QIP have been proposed \cite{Yoshimura2015creation, Nath2015hexagonal, Richerme2016two-dimensional} and experimentally explored \cite{Britton2012engineered, Garttner2017measuring, Seidelin2006microfabricated, Schmied2009optimal, Wang2020coherently, Donofrio2021radial, Xie2021open-endcap} with various technologies. 
For example, crystals of hundreds of ions in Penning traps, where the ion confinement is achieved with a static magnetic field and DC potentials, have been used to simulate two-dimensional Ising spin systems \cite{Britton2012engineered, Garttner2017measuring}. 
However, the ions rotate in the applied magnetic field of the Penning trap, creating additional challenges in optically addressing and measuring individual ion qubits.
Another approach is to minimize the impact of micromotion by carefully choosing the geometries of laser beams that are used to address ions in conventional RF traps.
Two-dimensional ion crystals have been studied in such systems for tens of ions \cite{Wang2020coherently, Donofrio2021radial, Xie2021open-endcap}, leading to experimental simulation of quantum magnetism \cite{Qiao2022observing}.
Alternatively, the long-range Coulomb-mediated spin-spin interactions in an ion chain can be tailored to effectively create a two-dimensional spin system \cite{Teoh2020machine,Korenblit2012quantum,Rajabi2019dynamical}.
However, such synthetic systems often require fine-tuned static or dynamic controls over the Hamiltonian, which can be experimentally costly and error-prone.

A way to obtain RF-micromotion-free and non-rotating two-dimensional ion system is to employ optical trapping of ions, using AC Stark shift from an optical beam. 
Indeed, optical tweezers \cite{Schneider2012influence, Schmidt2018optical} and optical lattices \cite{hoenig2023trapping} have been used to trap ions in one-dimension.
However, the typical depth of optical trapping potential is small compared to conventional RF and magnetic traps for experimentally feasible optical configurations.
Further, the optical potential is dependent on atomic states, and hence the lifetime of ions is limited by scattering from the optical beam into anti-trapping or weakly trapped atomic states. 
The lifetime can be increased by reducing the rate of scattering via increasing the frequency separation (`detuning') between the light and relevant atomic transitions.
For example, by changing the trapping beam from visible to near-infrared, the lifetime of an optically trapped Ba$^+$ ion was experimentally demonstrated \cite{Lambrecht2017long} to increase by a factor of 18 (from 166 ms to 3 s).
However, increased detuning also necessitates higher optical intensity to create a deeper trapping potential suitable for a multi-ion system.
A natural way to enhance the intensity of the optical field is to use a resonator cavity. 
In this manuscript, we study a hybrid trap architecture to trap 2D configuration of ions, where the in-plane confinement is provided by DC potential and the out-of-plane confinement is due to the optical standing wave in the cavity.

This manuscript is structured as follows. 
In Sec.~\ref{sec:optical 2D ion trapping}, we discuss the hybrid trapping potential created in our architecture.
We investigate the structural phase transition between 2D and 3D configurations of ions as a function of trap anisotropy in Sec.~\ref{sec:structural phase transitions}.
We also find that the transverse size of the optical cavity mode with respect to the characteristic size of the ion crystal plays an important role in the structural phase transitions.
Unlike in 1D, two-dimensional ion crystals may have metastable equilibrium configurations \cite{ransford2020high,Block2000crystalline}.
In Sec.~\ref{sec:potential barrier}, we estimate the potential barrier between the stable and metastable configurations for up to $N=9$ ions. 
We find that the potential barriers are higher than typical Doppler cooling temperatures.
This may explain the stability of observed configurations in experiments with 2D ion crystals \cite{Donofrio2021radial, Xie2021open-endcap, Block2000crystalline}.
In Sec.~\ref{sec:trapping lifetime and scalability}, we discuss the lifetime of multiple ions in the hybrid trap and identify the role of various heating mechanisms.
The 2D arrangement of ions readily allows the creation of 2D spin models via the standard M\o{}lmer-S\o{}rensen scheme \cite{Molmer1999multiparticle,Sorensen1999quantum, Monroe2021programmable}.
We discuss examples of spin-spin interactions between ions in 2D configurations, suitable for simulations of geometrically frustrated magnetic models in Sec.~\ref{sec:spinmodels}.
We also provide critical parameters (for $\mathrm{Yb}^+ $ ions) such as trap frequencies and scattering rate for experimentally achievable laser parameters and optical cavities for a given target radial trap frequency.


\section{optical 2D ion trapping}
\label{sec:optical 2D ion trapping}

\begin{figure}[ht]
    \centering
    \includegraphics[width=0.38\textwidth]{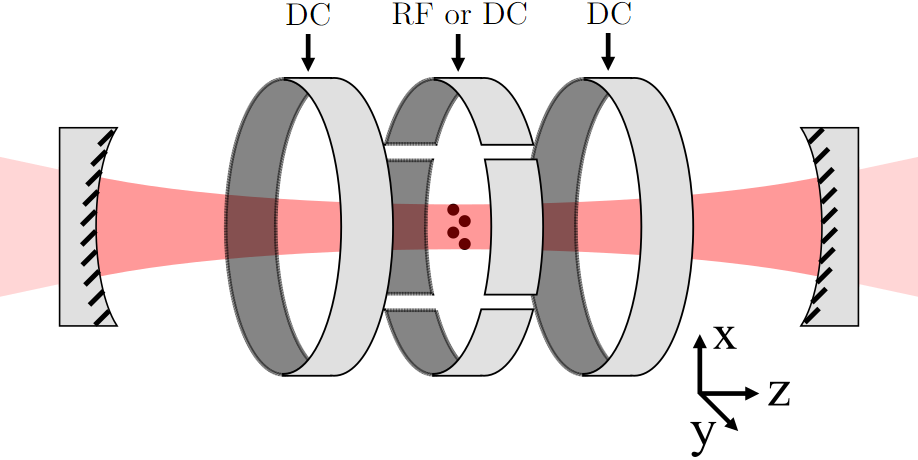}
    \caption{
    \textbf{2D optical cavity trapping setup.}
    Proposed 2D hybrid optical cavity trap setup (not to scale).
    Conventional trapping is provided by radio-frequency (RF) voltages on the central electrode and static (DC) voltages on the outer electrodes.
    Optical trapping along the $z$-axis is provided by AC Stark shift experienced by ions from the cavity beam standing wave.
    The hybrid trapping consists of $z$-trapping from the cavity beam and radial ($x,y$) trapping from DC voltages on the central electrode.
    }
    \label{fig:setup}
\end{figure}

The hybrid trapping architecture uses conventional trapping electrodes and a high-finesse optical cavity to trap 2D ion crystals, as schematically shown in Fig.~\ref{fig:setup}. 
At first, the ions are trapped in a 2D configuration with conventional DC and RF voltages \cite{Donofrio2021radial, Xie2021open-endcap}. 
Then, the ions are transferred to the micromotion-free optical trap by adiabatically ramping up the laser intensity in the cavity and the DC voltages (on the central ring electrode in Fig.~\ref{fig:setup}) while ramping down the RF potential and the DC voltages on the end electrodes.
The inhomogeneous spatial distribution of the laser intensity from the standing wave created inside an optical cavity results in a position-dependent AC Stark shift, providing confinement in the $z$ direction (i.e., out-of-plane for the 2D configuration). 
The trapping laser (with frequency $\omega_l$) can be either `blue' ($\omega_l>\omega_a$) or `red' ($\omega_l<\omega_a$) detuned with respect to the relevant atomic transitions ($\omega_a$), determining the sign of the AC Stark shift and whether the ions are trapped in a node or antinode of the standing wave in the cavity.

In the hybrid optical trapping regime, we consider the total potential of the $N$-ion system $U_\mathrm{total}$ consisting of three terms:
\begin{equation}
    U^\mathrm{total} \left( \{\mathbf{r}_i \}_{i=1}^N \right) = U^\mathrm{Coulomb} + U^\mathrm{DC} + U^\mathrm{opt},
\end{equation}
where $U^\mathrm{Coulomb}$ is the Coulomb potential, 
\begin{equation}
    U^{\mathrm{Coulomb}} \left( \{ \mathbf{r}_i \}_{i=1}^N \right)
    =\sum_{i<j} \frac{e^2}{ 4 \pi \varepsilon_0 \norm{\mathbf{r}_i - \mathbf{r}_j} }, 
\end{equation}
and $U^\mathrm{DC}$ is potential from the DC circular electrode, 
\begin{equation}
\begin{split}
    & U^{\mathrm{DC}} \left( \{ \mathbf{r}_i \}_{i=1}^N \right) \\
    &= \sum_i \ \frac{1}{2} m \Bigl[ \left(\omega_{x}^{\mathrm{DC}} x_i \right)^2
    + \left(\omega_{y}^{\mathrm{DC}} y_i \right)^2
    - \left(\omega_{z}^{\mathrm{DC}} z_i \right)^2 \Bigr]. 
\end{split}
\end{equation}
Here, $\mathbf{r}_i$ is the position vector of ion $i$, $m$ is the mass of ion, $\varepsilon_0$ is the permittivity of free space, and $\omega^{\mathrm{DC}}_{\xi}$ is the DC trap frequency, with $\xi \in \{x,y,z\}$. 
Here, We have assumed that the extent of the ion crystal in the radial direction is much smaller than the size of the DC electrodes, and hence the anharmonicity in $U^{\mathrm{DC}}$ can be neglected.
In addition, $ \omega_{z}^{\mathrm{DC}} $ should follow
$
    \left(\omega_{z}^{\mathrm{DC}} \right)^2  = \left(\omega_{x}^{\mathrm{DC}} \right)^2 + \left(\omega_{y}^{\mathrm{DC}} \right)^2
$
to satisfy Laplace's equation.  
The optical potential inside the cavity with a Gaussian trapping laser has the following form:
\begin{equation}
\label{eqn:optpot}
\begin{split}
    & U^\mathrm{opt} \left( \{ \mathbf{r}_i \}_{i=1}^N \right) = U_{\mathrm{depth}} ^{\mathrm{opt}} \times \\
    & \sum_i \left( \frac{w_0}{w(z_i)} \right)^2 \exp(\frac{-2(x_i^2+y_i^2)}{w^2(z_i)}) \sin^2 \left(\frac{2\pi z_i}{\lambda} \right), 
\end{split}
\end{equation}
where $w_0$ is the beam waist of Gaussian beam inside cavity, $w(z_i)$ is the beam radius at position $z_i$, $\lambda$ is the wavelength of the laser, and $U_{\mathrm{depth}} ^{\mathrm{opt}}$ is the optical trap depth, defined as the absolute value of the maximum AC Stark shift given by 
\begin{equation}
    U_{\mathrm{depth}} ^{\mathrm{opt}}
    = \abs{ -\sum_{a}\frac{3\pi c^2}{2\omega_a^3} \left( \frac{\Gamma_a}{\omega_a -\omega_l} + \frac{\Gamma_a}{\omega_a +\omega_l} \right) 
    I_{\mathrm{max}} }. 
\end{equation}
Here, $c$ is the speed of light, $\Gamma_a$ is the off-resonant scattering rate, $\omega_a $ is the atomic transition frequency between the ground and excited state $a$, and $I_{\mathrm{max}}$ is the maximum intensity inside the cavity. 
Ions are trapped at the plane of $z=0$ (see Fig.~\ref{fig:setup}), and hence the potential in the $z-$direction (in Eq.~\ref{eqn:optpot}) can be assumed to be harmonic as well.
We also assume that the center of the ion crystal is at the origin $x = y = 0$. 
In this paper, we focus on $ ^{171}\mathrm{Yb}^+ $ ions, which is both a popular choice for QIP experiments and suitable for longer lifetimes in a far-red-detuned optical trap (with the wavelength of the optical trapping beam at 1064 nm).

\begin{figure}[ht]
    \centering
    \includegraphics[width=0.45\textwidth]{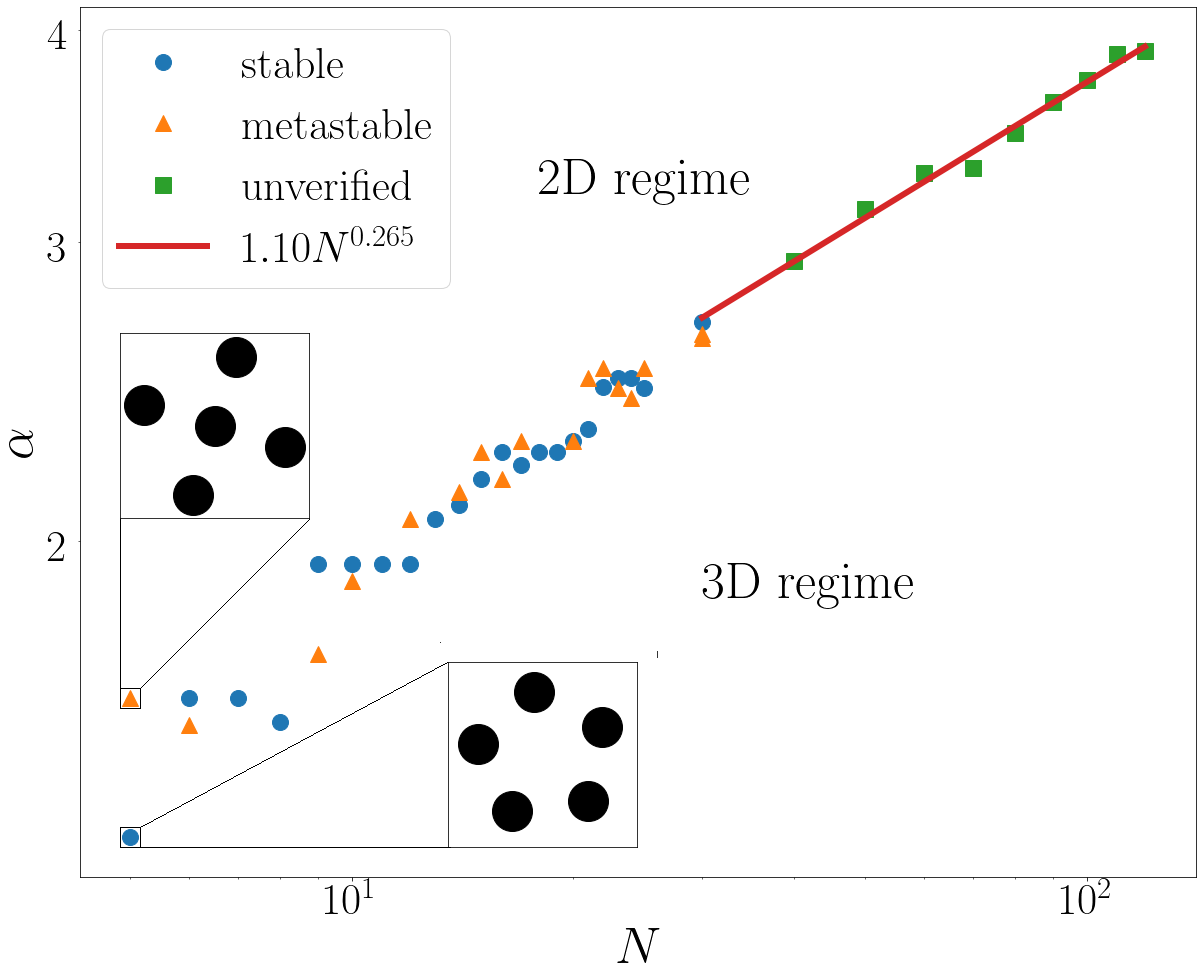} 
    \caption
    {
    \textbf{2D-3D structural phase transition in ion crystals.}
    Results are shown for a large cavity beam waist $ w_0 = 100 \ \mu \mathrm{m} $ compared to the size ($\approx 23 \ \mu \mathrm{m}$ for $N = 100$) of the $^{171}\mathrm{Yb}^+ $ ion crystal, with $ \omega_{r}^{\mathrm{DC}} / 2 \pi = 0.5 \ \mathrm{MHz} $.
    The determination of stable and metastable equilibrium positions are based on numerically calculates results (see main text).
    For small $ N \in [5,\ 30] $, we show the 2D-3D transition points with these stable (blue discs) and metastable (orange triangles) equilibrium positions. 
    For large $ N \in [40,\ 120] $, we do not verify whether the numerical equilibrium positions (green squares) are stable or metastable.
    The equilibrium points for $ N \in [30,\ 120] $ are fitted with a power law to obtain an approximate scaling of the 2D-3D phase transition points.
    (Inset) Stable and metastable equilibrium configurations for $N=5$.
    }
    \label{fig:transition_points}
\end{figure}

\begin{figure}[ht]
    \centering
    \includegraphics[width=0.45\textwidth]{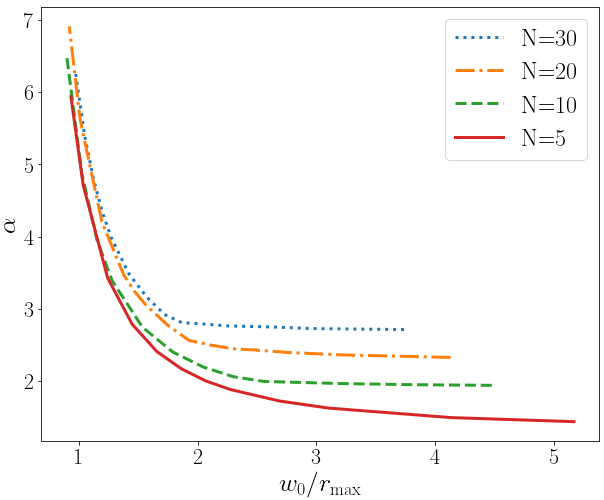} 
    \caption
    {
    \textbf{Structural phase transition points for ion crystals with different cavity beam waists.} The 2D-3D structural phase transition points for $ ^{171}\mathrm{Yb}^+ $ ions with $N = 5, \ 10, \ 20, \ 30$ (shown in different colors) versus beam waist $w_0$. The three curves are calculated using stable equilibrium positions. Ions are trapped in an optical cavity with $ \omega_{r}^{\mathrm{DC}} / 2 \pi = 0.5 \ \mathrm{MHz}$.
    }
    \label{fig:alpha_vs_w_0}
\end{figure}


\section{structural phase transitions}
\label{sec:structural phase transitions}

The anisotropy in trap frequencies determines whether a 2D ion crystal is stable against buckling into a 3D structure. 
We define the trap aspect ratio $\alpha \equiv \omega_z / \omega_r  $, where $ \omega_r \equiv \omega_x \approx \omega_y$, $\omega_z$ and $\omega_r$ are the trap frequencies at the origin.
Here, the trap frequencies arise from both DC and optical trapping.
The 2D-3D structural phase transition is analogous to the well-studied \cite{Dubin1993theory, Schiffer1993phase} `zigzag' phase transitions that arise in an ion chain. 
For $\alpha$ larger than the phase transition point, $\alpha_\mathrm{tr}$, the ion-crystal remains in the 2D phase.
In addition to the trap aspect ratio, the 2D-3D structural phase transitions in the optical cavity trap potential also depend on the cavity beam waist $w_0$. 
Note that the 3D phase may be unstable if the trap depth along $z-$direction is not adequate.

Figure~\ref{fig:transition_points} shows the critical trap aspect ratio $\alpha_\mathrm{tr}$ for various number of ions ($N=8$ to $N=120$), in the regime where the cavity waist $w_0$ is much larger than the size of the ion crystal $r_\mathrm{max}$.
We first calculate the equilibrium positions of the $N$-ion 2D crystal, by numerically optimizing the total potential $ U^\mathrm{total} \left( \{\mathbf{r}_i \}_{i=1}^N \right)$  with respect to $N$-ion positions.
In addition to the stable equilibrium configuration corresponding to the global potential minimum, there may exist metastable configurations corresponding to local minima as well \cite{Block2000crystalline}. 
However, we cannot verify if a given equilibrium configuration is the true global minimum point of the potential. 
Instead, in calculating various possible equilibrium positions, we re-run our algorithm tens of times, with different initial guesses for the ion positions.
If multiple equilibrium positions are obtained, we label the configuration with the minimum energy as `stable'.
Note that, it is possible that our algorithm misses to identify other equilibrium configurations, one of which could be the true stable configuration.
The 2D-3D structural phase transition points will in general be different for stable and metastable configurations, for the same $N$. 
We numerically find that the variations in $\alpha_\mathrm{tr}$ is $\lesssim 5\%$ for up to $N \le 25$, except for $N=$ 5, 9 and 21, where the variations are larger: $20\%$, $12\%$ and $7\%$ respectively.

The structural phase transition point for a given equilibrium configuration can be numerically determined by monitoring the lowest out-of-plane normal mode frequency, $\omega_\mathrm{lowest}^z$ as a function of $\omega_z$ while keeping the equilibrium positions of the 2D configuration fixed (i.e.~keeping $\omega_r$ fixed). 
As $\omega_z$ decreases, $\omega_\mathrm{lowest}^z$ decreases and eventually reaches zero.
Any further decrease in $\alpha $ results in an imaginary $\omega_\mathrm{lowest}^z$, suggesting a breakdown in the the normal mode approximation around the chosen equilibrium position.
The lowest normal mode frequency $\omega_\mathrm{lowest}^z=0$ corresponds to the 2D-3D structural phase transition for the given 2D (stable or metastable) configuration.
In Fig.~\ref{fig:transition_points}, we show the structural phase transition points for stable and metastable configurations for $N\le 30$.
However, numerically finding the true stable equilibrium configuration is challenging for larger number of ions.
Hence, for $N>30$, we show the transition point for an equilibrium configuration, without verifying whether that is a stable or metastable configuration.

The linearity of $\alpha_\mathrm{tr}$ versus $N$ in the log-log scaled plot of Fig.~\ref{fig:transition_points}, especially for large $N$ ($\gtrsim 30$), suggests that the transition points follow a power law, $ \alpha_\mathrm{tr} \approx 1.1 N^{0.265}$. 
Our numerically obtained power law exponent approximately agrees with the theoretical predictions in a harmonic potential for large $N$ (up to $ N=500 $), for which the exponents were calculated to be 0.25 \cite{Dubin1993theory} and 0.26 \cite{Schiffer1993phase}. 
Considerable deviation from this power law behaviour is observed, both in our numerics and the previous theoretical work \cite{Schiffer1993phase}, for small $N$.

When the cavity beam waist, $w_0$ is smaller or comparable to the size of the ion crystal, $r_\mathrm{max}$, ions at different locations experience substantially different local confinement.
The 2D-3D structural phase transition point is dependent on $w_0$ in this regime, as shown in Fig.~\ref{fig:alpha_vs_w_0}.
As $w_0/r_\mathrm{max}$ decreases, $\alpha$ scales up steeply and a much higher out-of-plane trap frequency (at the origin) is required to maintain sufficiently strong out-of-plane confinement for outer ions for the 2D phase to be stable.
For a given $N$, we find that $\alpha_\mathrm{tr}$ does not show a strong dependence on the beam waist beyond a given $w_0/r_\mathrm{max}$.
For example, $\alpha_\mathrm{tr}$ approaches its asymptotic value corresponding to very large $w_0/r_\mathrm{max}$ for $N = 10$ at $w_0/r_\mathrm{max}\approx2.5$.
Higher $N$ corresponds to a smaller $w_0/r_\mathrm{max}$ at which the asymptotic limit is achieved.
From a power efficiency point of view, it is practically beneficial to choose a cavity beam waist close to this value to maximize the optical trap depth for the available optical power, without being limited by the inhomogeneity in local optical confinement.
Note that, there may be additional considerations from the specific experimental protocol in choosing a cavity beam waist, such as maximum allowable differential AC Stark shift difference between inner and outer ions, and the desired structure of the normal modes.
With the study of structural phase transition combined with the trapping parameters illustrated in Sec.~\ref{sec:optical 2D ion trapping}, we find that it is feasible to trap ion crystals with several tens of ions in an optical cavity trap potential (see Appendix \ref{apdx:optical trapping parameters}).

\begin{figure}[ht]
    \centering
    \includegraphics[width=0.45\textwidth]{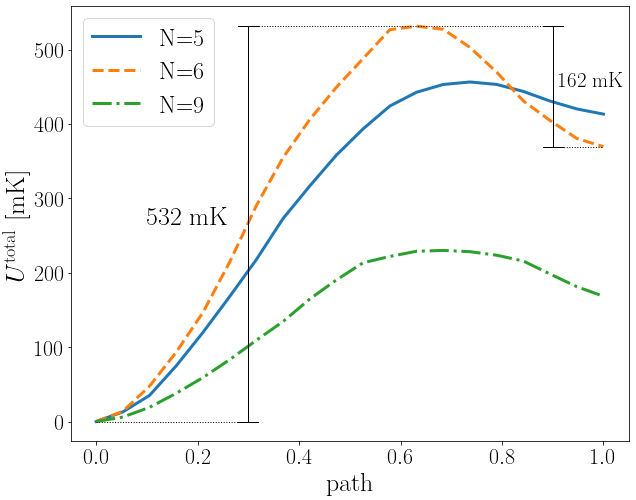} 
    \caption
    {\textbf{Potential barrier between equilibrium configurations.} 
     Numerically optimized paths between the two equilibrium configurations for various $N$ (see Appendix \ref{apdx:optimization algorithm} for the algorithm).
     On the horizontal axis, $\mathrm{path} = 0$ and 1 represent the stable and metastable equilibrium configurations, respectively. 
     The potential energy of the stable equilibrium configurations of $N = 5, \ 6, \mathrm{and} \ 9 $ are set to zero for comparison purposes, but the three paths belong to different configuration spaces. 
     The two potential barriers for $N = 6$ stable and metastable configurations (532 mK and 162 mK, respectively) are labelled.
     For $N = 7$ and 8, we find only one 2D equilibrium configuration. }
    \label{fig:pe_optimized}
\end{figure}


\section{potential barrier between equilibrium positions}
\label{sec:potential barrier}
Trapped ions experience various heating mechanisms, which can lead to a change in their equilibrium configuration, where multiple 2D equilibrium configurations exist.
QIP experiments generally rely on optically addressing individual ions and exciting motional modes \cite{Molmer1999multiparticle, Sorensen1999quantum} that are specific to a given equilibrium configuration, and hence a change in the ion configuration may be catastrophic.
In this section, we numerically investigate the stability of 2D ion crystals by studying the potential barrier between equilibrium configurations, which quantifies the minimum kinetic energy required to transition between them.

Assuming that the out-of-plane trap frequency is strong enough such that ions can only move in the $z=0$ plane, each $N$-ion position is represented by a point in a $2N$-dimensional configuration space.
When the $N$-ion system transitions from one equilibrium configuration to another, it traces a continuous path connecting the two equilibrium positions in the same space.
We denote the maximum potential energy along the path as the peak potential for that path.
Among infinitely many possible paths, we define the difference between the smallest peak potential and $ U^\mathrm{total}$ of a given equilibrium configuration as its corresponding potential barrier.

To identify the path with the smallest peak potential and obtain upper bounds of potential barriers for a given $N$, we use an optimization algorithm discussed in detail in Appendix \ref{apdx:optimization algorithm}. 
In Fig.~\ref{fig:pe_optimized}, we show the results obtained using this optimization algorithm for 2D ion crystals with $N = 5, \ 6, \mathrm{and} \ 9 $. 
We find only one metastable equilibrium configuration for each $N$. 
For $N = 5 \ \mathrm{and} \ 9 $, the potential barriers corresponding to metastable equilibrium configurations (the right end of the plot) are around $ 40$ to $60 \ \mathrm{mK}$, while the barrier for the stable configuration are $456\;\mathrm{mK}$ and $230\;\mathrm{mK}$, respectively. 
For $N=6$, the barrier for the metastable state is relatively higher ($162 \ \mathrm{mK}$), while the barrier for the stable state ($532\;\mathrm{mK}$) is comparable to $N=5$ and 9. 
We note that these values for the potential barriers are higher than the typical Doppler cooling temperatures and hence we do not expect to see fluctuations between these stable and metastable configurations, unless we provide extra energy into the system (e.g., during a non-adiabatic transfer from the conventional trap into the hybrid trap).
The potential barrier corresponding to metastable configuration can be comparable to the optical trap depth (can achieve at least $\sim 100$ mK in the optical cavity trap, see Table \ref{tab:trap_parameters} ) and hence the stability of such configurations may be vulnerable to external perturbations. 
In that case, it will be desirable to obtain the stable 2D configuration in the conventional trap before transferring to the hybrid optical trap.


\section{trapping lifetime and scalability}
\label{sec:trapping lifetime and scalability}

Compared to conventional potentials, optical potentials generally result in a significantly shorter ion-trapping lifetime \cite{Schmidt2018optical,Lambrecht2017long,Schneider2012influence}.
Nevertheless, to be practically valuable, this platform requires optical trapping lifetimes that extend to at least over the same timescale as quantum experiments involving multiple ions (preferably 100 ms or more) \cite{Monroe2021programmable}.
Optical trapping lifetime is defined as the time it takes for the optical trapping probability $p_{\mathrm{opt}}$ to decrease to $1/e$, where $p_{\mathrm{opt}}$ is the probability for retaining all ions in the trap.

We find that scattering to anti-trapping atomic states of the ion is the limiting factor for the optical trapping lifetime, consistent with the observations of prior optical trapping experiments \cite{Lambrecht2017long}. 
For example, the metastable $^{2}D_{3/2} $ manifold of $ ^{171}\mathrm{Yb}^+ $ ion experiences a positive AC Stark shift interacting with the $1064 \ \mathrm{nm}$ Gaussian laser, and hence the $^{2}D_{3/2} $ manifold is anti-trapping.
This manifold also has a relatively long atomic lifetime of $61.8 \ \mathrm{ms}$ \cite{Schacht2015hyperfine}, over which ions can escape the trap. 
Since the trapping probability $p_{\mathrm{opt}}$ is defined for the $N$-ion system, the trapping probability for each ion at the end of the trapping lifetime should be $(1/e)^{1/N} $ such that $p_{\mathrm{opt}}(\tau) = 1/e$. 
If the scattering to metastable states is the only loss mechanism, we have 
\begin{align}
    e^{- \Gamma_\mathrm{meta} \tau} &= \left( \frac{1}{e} \right) ^{1/N} \\
\label{tau}
    \tau &= \frac{1}{\Gamma_\mathrm{meta} N },   
\end{align}
where $ \tau $ is the optical trapping lifetime and $ \Gamma_\mathrm{meta} $ is the scattering rate to metastable states. 
For example, if the total off-resonant scattering rate for $ ^{171}\mathrm{Yb}^+ $ is $ \Gamma_\mathrm{off} = 30 \ \mathrm{s^{-1}} $, then $ \Gamma_\mathrm{meta} = 0.15 \ \mathrm{s^{-1}} $, since the branching ratio between $^{2}P_{1/2} \rightarrow {}^{2}S_{1/2} $ and $^{2}P_{1/2} \rightarrow {}^{2}D_{3/2} $ is $ 200:1$ \cite{olmschenk2007manipulation}. 
For a 2D ion crystal with $ N = 20 $ $ ^{171}\mathrm{Yb}^+ $ ions, the scattering rate limited optical trapping lifetime is $ \tau=333 \ \mathrm{ms} $.
The loss due to scattering to metastable states becomes dominant for large $N$, providing an upper bound of $ (\Gamma_\mathrm{meta} N)^{-1} $ for optical trapping lifetime.  
If we apply a repumping laser actively transferring the population out of the metastable states, the optical trapping lifetime can be extended beyond this upper bound. 
For example, in Lambrecht et al.~\cite{Lambrecht2017long}, the optical trapping lifetime of a single $ \mathrm{Ba}^+ $ ion was increased from 21 ms to 166 ms by applying repumping lasers, which is an increase of nearly 8 times. 
In the above analysis, we assume that each ion-loss event is independent from others and all ions have the same scattering rates.

In addition, collisions with background gas particles and recoil effect from scattering can also heat up the system, although these effects are relatively minor (see Appendix \ref{apdx:heating effects}).


\section{Vibrational modes and spin-spin interactions}
\label{sec:spinmodels}

\begin{figure*}[ht]
    \centering
    \includegraphics[width=0.8\textwidth]{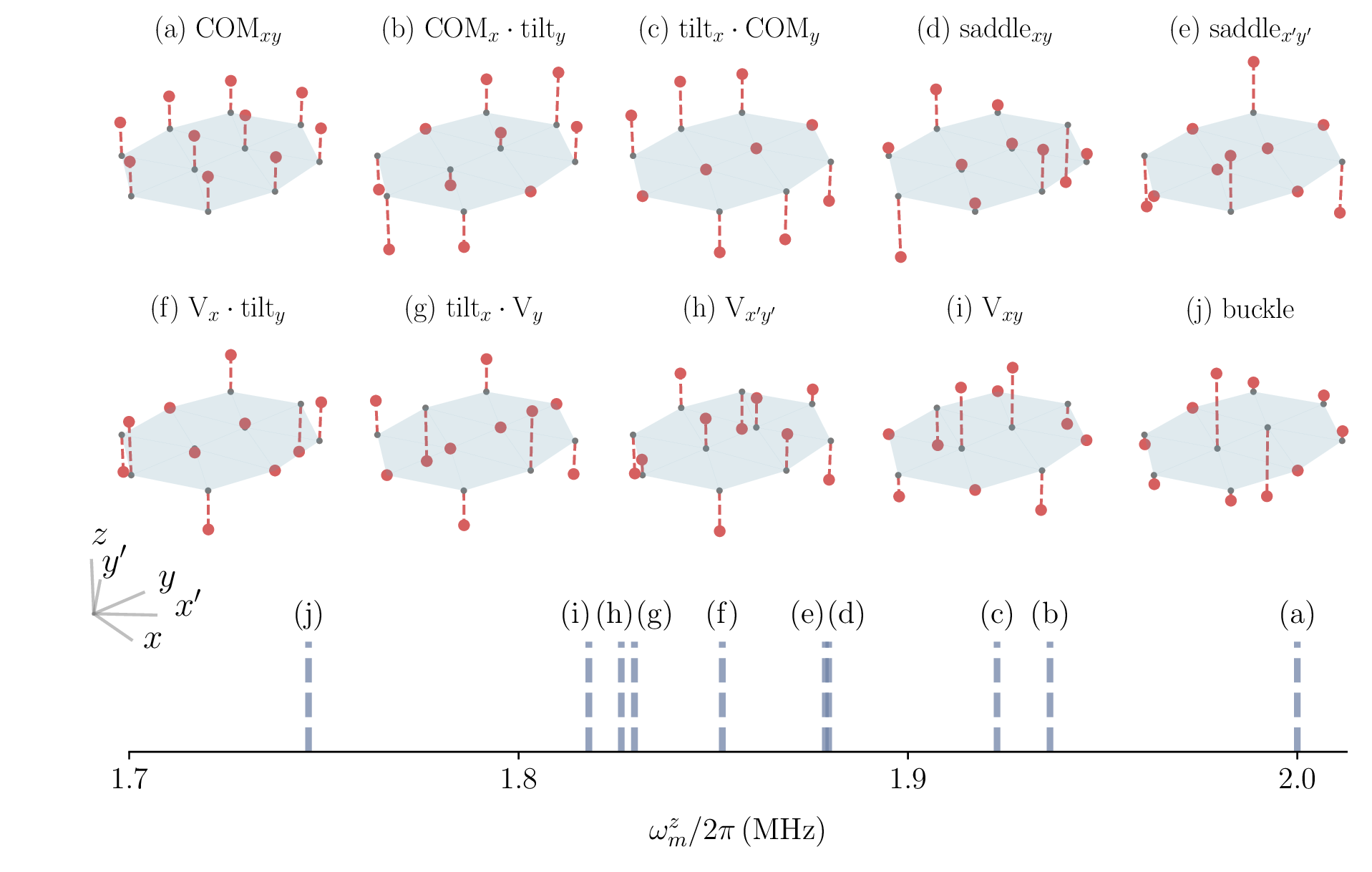}
    \caption{
    \textbf{Out-of-plane ($z$) vibrational modes of an $N= 10$ 2D ion crystal.}
    The trap strengths are $\omega_{r} / 2\pi \approx 0.5 \, \mathrm{MHz}$ and $\omega_z / 2\pi = 2 \, \mathrm{MHz}$ for these numerical calculations.
    The mode labels are described in the text.
    To avoid degeneracies (see main text) arising from rotational symmetry in the total potential, $U^{\mathrm{total}}$, we have introduced a $10 \, \%$ anisotropy in the trap frequencies along $x$ versus $y$-directions.    
    }
    \label{fig:2dtransversemodes}
\end{figure*}

2D ion crystals exhibit a rich vibrational mode structure.
The highest frequency mode along the out-of-plane direction is the center of mass (COM) mode, with the next highest frequency modes typically being the two tilt modes (see Fig.~\ref{fig:2dtransversemodes}).
Furthermore, Fig.~\ref{fig:2dtransversemodes} shows that the modes of the $N=10$ 2D ion crystal have composite structure comprised of the typical modes in a 1D chain, i.e.~the COM mode, tilt mode and `V'-shaped mode.
For example, the saddle$_{xy}$ mode is a composite mode of tilt$_{x}$ $\cdot$ tilt$_{y}$.
Additionally, we observe that the outer ions are the dominant participants in the higher modes and the inner ions are the dominant participant in the lower modes, with an exception being the COM mode where all ions participate equally.
As there are more outer ions than inner ions, more energy is required to move the ensemble of outer ions compared to the ensemble of inner ions.

The equilibrium configuration of an ion crystal violates the underlying symmetry of a trap that is isotropic along $x$ and $y$-directions (this is an example of spontaneous symmetry breaking).
However, We find that multiple modes, e.g.~the tilt modes, are still approximately degenerate for an isotropic trap.
By introducing slight anisotropy $(\sim 8 \, \%)$, the degeneracy in the tilt modes can be broken by approximately $0.5 \, \%$, however, a greater anisotropy $(\gtrsim 20 \, \%)$ is needed to break the degeneracy of the saddle modes by a similar amount.

In the above calculations, we assumed that the cavity beam waist is much larger than the characteristic size of the ion crystal.
For smaller cavity beam waists, there will be inhomogeneous trapping of the inner and outer ions, stronger on the inner ions and weaker on the outer ions.
Consequently, the frequency and shape of the modes will change \cite{teoh2021manipulating}.
For example, the COM mode (with equal amplitude on all ions) will not be an exact eigenmode for a finite cavity beam waist.
However, it is still an approximate mode.
As an example, for $w_0 = 21 \ \mu \mathrm{m}$ and $r_{\mathrm{max}} \approx 7.8 \ \mu \mathrm{m}$ (from Appendix~\ref{apdx:optical trapping parameters}), the relative amplitudes of the weakest and strongest participating ions is approximately 0.56.
Increasing the cavity beam waist will reduce the imbalance between the amplitudes of the participating ions in the COM mode \cite{teoh2021manipulating}.

As mentioned in Sec.~\ref{sec:structural phase transitions}, the lowest frequency mode plays an important role when characterizing the 2D-3D phase transition.
Analogous to the zigzag phase in a 1D-2D phase transition, the 3D configuration resembles the eigenvector of the lowest frequency mode, i.e.~the middle ions buckle out of the 2D plane past the phase transition point.

The rich vibrational mode structure can be used in generating interesting spin-spin interaction profiles for the simulation of quantum spin models \cite{Monroe2021programmable, Blatt2012quantum}. 
By applying suitable spin-dependent forces (SDF), generated from laser beams, spin-spin interactions can be engineered with controls (in principle arbitrary) over strength, range, and sign of the interaction profile.
Spin models such as long-range Ising and XY have been studied experimentally in a 1D chain in conventional RF traps \cite{Kim2009entanglement,Korenblit2012quantum,Teoh2020machine, Kotibhaskar2023programmable} and also in 2D rotating crystals in Penning traps.
Extending these protocol to 2D non-rotating crystals in our hybrid trap will enable access to more complex quantum simulations of spin models, such as of 2D frustrated Hamiltonians.

\begin{figure*}[ht]
    \centering
    \includegraphics[width=1\textwidth]{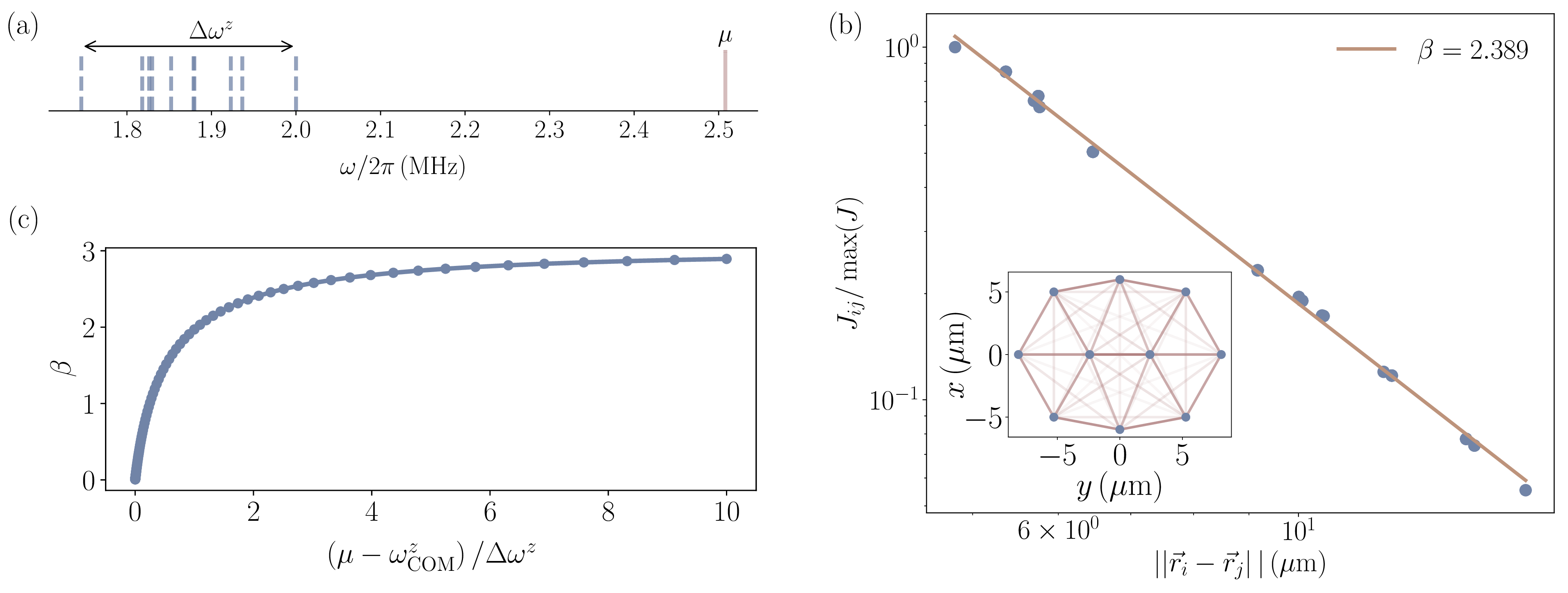}
    \caption{
    \textbf{Frustrated spin-spin interactions for $N=10$ ions.}
    (a) Out-of-plane ($z$) mode frequencies $\omega^{z}_{m}$ (blue dashed) alongside the SDF frequency $\mu$ (red solid).
    (b) Spin-spin interactions $J_{ij}$ (calculated from Eq.~\ref{eqn:Jij}) follow a spatial power-law decay with exponent $\beta$ (Eq.~\ref{eqn:powerlawspingraph}).    
    (Inset) the resulting interaction graph.
    All the interactions are anti-ferromagnetic for this $\mu$.
    The strongest interactions form a triangular lattice, creating a frustrated spin system.
    (c) Variation in $\beta$ versus the SDF frequency.
    }
    \label{fig:2dspinpowerlaw}
\end{figure*}

In the commonly used M\o{}lmer-S\o{}rensen scheme \cite{Molmer1999multiparticle,Sorensen1999quantum}, the effective Hamiltonian $\hat{H}$ of the trapped ion system, when driven by multiple SDFs, is:
\begin{equation}
    \hat{H} = \sum_{i<j} J_{ij} \hat{\sigma}^{(x)}_i \hat{\sigma}^{(x)}_j,
\end{equation}
where,
\begin{equation}
    J_{ij} = E_{\mathrm{recoil}} \sum_{n} \Omega_{in} \Omega_{jn} \sum_{m} 
    \frac{b_{im} b_{jm}}{\mu_n^2 - \omega_m^2}.
    \label{eqn:Jij}
\end{equation}

Here, $J_{ij}$ is the spin-spin interaction strength between ions $i$ and $j$, $E_{\mathrm{recoil}}$ is the characteristic recoil energy from the laser field driving the SDFs, $\mu_n$ is the $n^{\mathrm{th}}$ SDF frequency, $\omega_m$ is the $m^{\mathrm{th}}$ normal mode frequency and $\Omega_{in}$ is the Rabi frequency of ion $i$ for the $n^\mathrm{th}$ SDF.

Like a chain of ions, a power law interaction profile can be generated from a single SDF \cite{monroe2015quantum}.
The form of the interaction is,
\begin{equation}
    J_{ij} \propto  \frac{1}{\lVert \mathbf{r}_i - \mathbf{r}_j \rVert^{\beta}},
    \label{eqn:powerlawspingraph}
\end{equation}
where, $\beta$ is the power law exponent.
Figure \ref{fig:2dspinpowerlaw} shows that the exponent $\beta$ can be tuned between 0 and 3 for a 2D configuration, when the SDF frequency is varied.
A power law like interaction in the 2D configuration, in comparison with a 1D chain, exhibits magnetic frustration naturally, due to the additional spatial degree of freedom and the configuration of the ions (shown in Fig.~\ref{fig:2dspinpowerlaw}) resembling a triangular lattice. 

Other lattice structures can be obtained via optical engineering of the Raman beams, similar to the case of 1D ion chains \cite{Korenblit2012quantum,Teoh2020machine}. 
Alternatively, the triangular lattice can be further mapped to a square, Kagome or other lattice structures by augmenting the analog simulation with digital gates \cite{Rajabi2019dynamical}.


\section{Conclusion}
\label{sec:discuss}
To summarize, our study demonstrates the potential of utilizing an optical cavity to trap 2D ion crystals, offering enhanced trap depth and extended lifetime when compared to existing tweezer-based optical ion trapping experiments.
We estimate that the off-resonant scattering to anti-trapping states from the cavity beam, rather than heating due to photon recoil and collisions, is the limiting factor for trapping lifetime, particularly for large $N$.
To increase the optical trapping lifetime, one can either reduce the total off-resonant scattering rate or apply a repumping laser. 
An alternative approach can be to use ion species with a smaller atomic number, which do not have any metastable states between the lowest atomic $S$ and $P$ manifolds.
However, there may be additional challenges when using such ions, such as reduced AC Stark shift at the same optical intensity.

In addition, ions may be lost during the transferring process between conventional and optical trapping regimes, if the transfer is not adiabatic enough so as to provide the ions with energy to overcome the optical trap depth.
We may also want to minimize any motional excitation to avoid the necessity to cool the ions in the optical trap, as cooling protocols may populate anti-trapping states.
Due to the shallow optical trap depth, detection of quantum states involving ion fluorescence is likely unfeasible without losing the ion from the trap.
Thus, transferring the ions back into the conventional trap is preferable.

For QIP experiments with this system, the quantum coherence will be destroyed by all off-resonant scattering events, including those that do not lead to ion loss from the trap.
Therefore, coherence time of the full system is typically much shorter than the optical trapping lifetime.
The scattering rate-limited coherence time scales as $1/(\Gamma_\mathrm{off} N)$, where $\Gamma_\mathrm{off}$ is the total off-resonant scattering rate. 
Reducing the scattering rate $\Gamma_\mathrm{off}$ is the only way to overcome this fundamental limitation, leading to our choice of 1064 nm laser for this proposal. 
Alternatively, blue detuned trapping lasers, where ions are trapped at the intensity minimum, can be investigated \cite{gerbier2010heating} for reducing scattering rates that limit the coherence time and trapping lifetime.
However, as most ion species has relevant transitions in the ultraviolet regime, it may be challenging to create a deep optical potential with blue detuned lasers.
The suitability of red vs blue detuned trapping will also depend on the atomic structure of the species.
For example, $\mathrm{Ba}^+ $ and $ \mathrm{Yb}^+ $ ions have very different branching ratios into metastable $D$-states, such that under the same $\Gamma_\mathrm{off}$, $\Gamma_\mathrm{meta}$ for $ \mathrm{Yb}^+ $ is much lower than $ \mathrm{Ba}^+ $. 
We think that the low $\Gamma_\mathrm{meta}$ makes $ \mathrm{Yb}^+ $ a better candidate compared to $ \mathrm{Ba}^+ $ for red detuned optical traps, but $ \mathrm{Ba}^+ $ can be a potential candidate for blue detuned optical traps.


\acknowledgments
We acknowledge financial support from the Canada First Research Excellence Fund (CFREF), the Natural Sciences and Engineering Research Council of Canada (NSERC) Discovery program (RGPIN-2018-05250), University of Waterloo, and Innovation, Science and Economic Development Canada (ISED).


\bibliography{ref}


\appendix

\section{2D Optical Trapping Parameters}
\label{apdx:optical trapping parameters}

\begin{table*}[ht]
\centering
\begin{tabular}{|c|cccc|}
\hline
Ion number $N$           & \multicolumn{1}{c|}{5} & \multicolumn{1}{c|}{10} & \multicolumn{1}{c|}{20} & 30 \\ \hline
Ion configuration        & \multicolumn{1}{c|}{ [5] } & \multicolumn{1}{c|}{ [2, 8] } & \multicolumn{1}{c|}{ [1, 7, 12] } & [5, 10, 15] \\ \hline
Radial DC trap frequency $ \omega_{r}^{\mathrm{DC}} $ & \multicolumn{4}{c|}{$ 2\pi \times 0.5 $ MHz}                          \\ \hline
Laser wavelength $\lambda$ & \multicolumn{4}{c|}{1064 nm}                          \\ \hline
Minimum ion spacing [$\mu$m] & \multicolumn{1}{c|}{ 5.7 } & \multicolumn{1}{c|}{ 4.7 } & \multicolumn{1}{c|}{ 4.8 } & 4.4 \\ \hline
Ion crystal radius [$\mu$m] & \multicolumn{1}{c|}{ 4.8 }  & \multicolumn{1}{c|}{ 7.8 }  & \multicolumn{1}{c|}{ 10.9 }   & 13.4 \\ \hline
Trapping beam waist $w_0$ [$\mu$m] & \multicolumn{1}{c|}{ 14.4 }  & \multicolumn{1}{c|}{ 21.0 }  & \multicolumn{1}{c|}{ 27.3 }   & 26.8 \\ \hline
Minimum required AC Stark shift at center [$\mathrm{MHz \cdot h}$] ([mK]) & \multicolumn{1}{c|}{ 287 (13.8)} & \multicolumn{1}{c|}{ 361 (17.3)} & \multicolumn{1}{c|}{ 475 (22.8)} & 600 (28.8)\\ \hline
Differential AC Stark shift at center [$\mathrm{kHz \cdot h}$] & \multicolumn{1}{c|}{ 15 } & \multicolumn{1}{c|}{ 19 } & \multicolumn{1}{c|}{ 26 } & 32 \\ \hline
Differential AC Stark shift at edge [$\mathrm{kHz \cdot h}$] & \multicolumn{1}{c|}{ 7.7 } & \multicolumn{1}{c|}{ 9.0 } & \multicolumn{1}{c|}{ 12 } & 12 \\ \hline
Minimum required cavity intensity at center [$\mathrm{W/m^2}$] & \multicolumn{1}{c|}{$ 9.22 \times 10^{11} $} & \multicolumn{1}{c|}{$ 1.16 \times 10^{12} $} &  \multicolumn{1}{c|}{$ 1.53 \times 10^{12} $} & $ 1.93 \times 10^{12} $  \\ \hline
Cavity finesse $\mathcal{F}$ & \multicolumn{4}{c|}{3000}                             \\ \hline
Minimum required laser power [W]  & \multicolumn{1}{c|}{ 0.31  }  & \multicolumn{1}{c|}{ 0.84 }  & \multicolumn{1}{c|}{ 1.9 }   & 2.3 \\ \hline
Off-resonant scattering rate of an ion at center $\Gamma_\mathrm{off}$ [$\mathrm{s^{-1}} $] & \multicolumn{1}{c|}{ 2.8 }  & \multicolumn{1}{c|}{ 3.6 } & \multicolumn{1}{c|}{ 4.7 }   & 5.9  \\ \hline
\end{tabular}
\caption
{\textbf{Optical trapping parameters for $ ^{171}\mathrm{Yb}^+ $.}
The ion configuration [$n_1,n_2,...$] indicates that $n_1$ ions populate the `innermost ring', $n_2$ ions the next inner ring, and so forth. 
The trapping beam waist is determined from the ion crystal radius based on results from Fig.~\ref{fig:alpha_vs_w_0}.
For each $N$, the parameters are determined for the stable equilibrium position of the ion crystal.
The AC Stark shift is the energy shift of the ground $^{2}S_{1/2}$ state.
The differential shift is the difference between AC Stark shifts of two hyperfine states of $6 \, ^{2}S_{1/2}$ manifold $\ket{F=0, m_F=0}$ and $\ket{F=1, m_F=0}$, with a $\pi$-polarized laser, and the center refers to the center of equilibrium position of ion crystal, whereas the edge is the position of outer-most ion. 
The minimum required optical power of the incident laser is calculated based on a cavity finesse of 3000.
The scattering rate refers to the total off-resonant scattering rate of one ion under the minimum cavity intensity at the center of the ion crystal.
} 
\label{tab:trap_parameters}
\end{table*}

Table.~\ref{tab:trap_parameters} lists the minimum required trapping parameters for 2D ion crystals with different $N$ to maintain the 2D regime. 
Radial DC trap frequency determines the 2D ion crystal size and ion spacing, which is chosen such that the ions can be optically addressed for QIP experiments with low crosstalk between them \cite{motlakunta2023preserving}.

Since AC Stark shift and off-resonant scattering rate are proportional to intensity, the listed parameters help to reveal the proportionality constants. 
The listed minimum required laser power is less than what is practically achievable for 1064 nm laser and the damage threshold for cavity mirror coatings.
Hence, the optical trap depth can be significantly improved beyond the listed minimum AC Stark shift.


\section{Optimization Algorithm for Potential Barrier}
\label{apdx:optimization algorithm}

\begin{figure}[ht]
    \centering
    \includegraphics[width=0.4\textwidth]{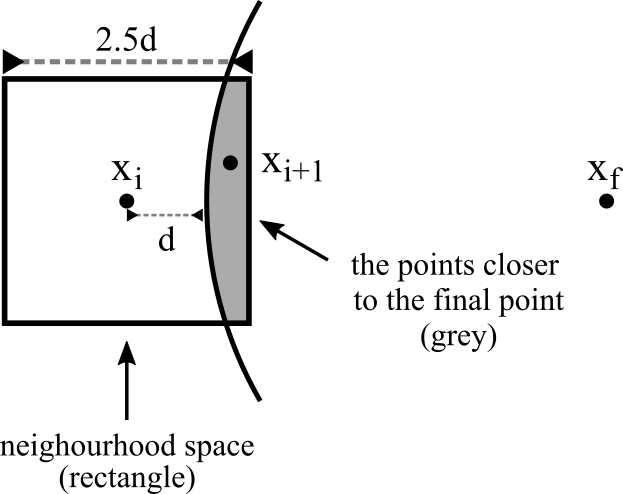} 
    \caption{
    \textbf{Schematic diagram of one iteration in the optimization algorithm.}
    }
    \label{fig:algorithm}
\end{figure}

In Sec.~\ref{sec:potential barrier}, potential barriers between stable and metastable equilibrium positions of 2D ion crystal is studied and presented. 
The following optimization algorithm is used to identify the path with the smallest peak potential. 
This algorithm has the following steps: 
\begin{enumerate}
    \item For a 2D $N$-ion crystal, the stable equilibrium position is set to be the initial point $\mathbf{x}_i$ for $i = 0$, and a metastable equilibrium position is set to be the final point $\mathbf{x}_f$. 
    The points are defined in the $2N$-dimensional configuration space. 
    \item Define a neighbourhood space of $\mathbf{x}_i$ with linear size $\varepsilon$, as shown in Fig.~\ref{fig:algorithm}, which contains a set of $N$-ion positions that are relatively close to the initial point. 
    \item Select all points in the neighbourhood space that are closer to the final equilibrium position by a distance $d$ or larger, which is the grey region in Fig.~\ref{fig:algorithm}. 
    Since the size of the neighbourhood space scales up exponentially with respect to $N$, numerically selecting all the points is not feasible. 
    Instead, $n_r=1000$ number of randomly sampled points in the grey region are selected.
    We choose $d \approx \left\lVert \mathbf{x}_0 - \mathbf{x}_f \right\rVert / 20$ and neighbourhood space linear size $\varepsilon=2.5 d$.
    \item Calculate the potential energy of all the selected points in the neighbourhood space. 
    \item Assign each selected point with a transition probability according to the Boltzmann probability distribution
    \begin{equation}
        p_{j} = \mathrm{normalize}\left[ e^{ - (E_{j} - E_{i}) / k_B T_\mathrm{p} } \right], 
    \end{equation}
    where $E_i$ is the potential energy of the initial point, $E_j$ is for the selected point, $k_B$ is the Boltzmann constant.
    $T_\mathrm{p}$ is a hyper-parameter with units of temperature, which determines the `volume of the trajectory space' under consideration.
    We empirically choose $T_\mathrm{p} = 1$ mK, comparable to the temperature of a Doppler cooled ion.
    \item Take a sample (denoted as $x_{i+1}$) from the selected points according to the transition probability $p_j$.
    
\end{enumerate}
The sampled $x_{i+1}$ in step 6 will replace the initial point $x_i$ in step 2, and by repeating steps 2$-$6 until the distance between the sampled point and the final point is smaller than $d$.
A path connecting the initial and final equilibrium positions can be obtained using this algorithm. 
This algorithm biases the trajectories towards smaller potential peaks.
We calculate 10 paths with this approach and take the smallest potential peak to find an upper bound of the potential barrier.
Figure~\ref{fig:pe_optimized} plots the trajectory with the smallest potential peak for a given $N$.


\section{Heating Effects}
\label{apdx:heating effects}
The theoretical estimation of optical trapping lifetime is based on the analysis of ion scattering rate and different heating mechanisms. 
In Sec.~\ref{sec:trapping lifetime and scalability}, we discussed how scattering to anti-trapping states can be the limiting factor in optical trapping lifetime. 
This conclusion is supported by quantitative analysis of other sources of heating, which include collisions with background gas particles and recoil heating from photon scattering. 

Since the ion trap chamber is not a perfect vacuum, ions can collide with background gas particles (mostly hydrogen molecules). 
Using the Langevin collision model, we find that the Langevin collision rate of a single ion with hydrogen molecules is only 1.3 per hour, for a pressure of $10^{-11} \ \mathrm{mbar}$ at the room temperature. 
Although a Langevin collision event can provide enough energy for the ion to escape the trap, the collision rate is negligibly small compared to typical optical trapping lifetime. 
There are non-Langevin collisions, with impact parameter larger than a critical value, such that the gas particles are not `captured' by the ion. 
Significantly smaller energies are exchanged between the two particles after the collision. 
The non-Langevin collision rate is also relatively small compared to the expected optical trapping lifetime. 
Our calculations show that the heating rate caused by non-Langevin collisions with hydrogen molecules is less than $ 0.1 \ \mathrm{mK/s}$ for each ion, under $ 10^{-11} \ \mathrm{mbar} $ pressure and $ 300 \ \mathrm{K} $ temperature. 

Recoil heating caused by the photon scattering from the optical cavity trapping beam is even less impactful.  
For a $ ^{171}\mathrm{Yb}^+ $ ion interacting with $ \lambda = 1064 \ \mathrm{nm}$ laser, the recoil energy is $ E_{\mathrm{rec}} \approx 5\times 10^{-5}\ \mathrm{mK} \cdot k_B$ per ion per scattering event. 
Since $ 1064 \ \mathrm{nm} $ is a far detuned laser for $ ^{171}\mathrm{Yb}^+ $ ion such that off-resonant scattering rate is suppressed (see Table.~\ref{tab:trap_parameters}), this recoil heating rate should have negligible impact on the optical trapping lifetime. 

\end{document}